\documentclass[preprint2]{aastex62}

\usepackage{hyperref}
\usepackage{amsmath}
\usepackage{booktabs}
\usepackage{graphics}
\usepackage{changepage}
\usepackage{ulem}

\begin{document}

\title{Mass--Radius relationship for M dwarf exoplanets: Comparing nonparametric and parametric methods}
\author[0000-0001-8401-4300]{Shubham Kanodia}
\affiliation{Department of Astronomy \& Astrophysics, The Pennsylvania State University, 525 Davey Laboratory, University Park, PA 16802, USA}
\affiliation{Center for Exoplanets and Habitable Worlds, The Pennsylvania State University, 525 Davey Laboratory, University Park, PA 16802, USA}

\author[0000-0003-2862-6278]{Angie Wolfgang}
\affiliation{Department of Astronomy \& Astrophysics, The Pennsylvania State University, 525 Davey Laboratory, University Park, PA 16802, USA}
\affiliation{Center for Exoplanets and Habitable Worlds, The Pennsylvania State University, 525 Davey Laboratory, University Park, PA 16802, USA}
\affiliation{NSF Astronomy and Astrophysics Postdoctoral Fellow}

\author[0000-0001-7409-5688]{Gudmundur K. Stefansson}
\affiliation{Department of Astronomy \& Astrophysics, The Pennsylvania State University, 525 Davey Laboratory, University Park, PA 16802, USA}
\affiliation{Center for Exoplanets and Habitable Worlds, The Pennsylvania State University, 525 Davey Laboratory, University Park, PA 16802, USA}
\affiliation{NASA Earth and Space Science Fellow}

\author[0000-0001-5256-6418]{Bo Ning}
\affiliation{Department of Statistics and Data Science, Yale University, 24 Hillhouse Avenue, New Haven, CT 06511, USA}

\author[0000-0001-9596-7983]{Suvrath Mahadevan}
\affiliation{Department of Astronomy \& Astrophysics, The Pennsylvania State University, 525 Davey Laboratory, University Park, PA 16802, USA}
\affiliation{Center for Exoplanets and Habitable Worlds, The Pennsylvania State University, 525 Davey Laboratory, University Park, PA 16802, USA}

\correspondingauthor{Shubham Kanodia}
\email{szk381@psu.edu}

\begin{abstract}
Though they are the most abundant stars in the Galaxy, M dwarfs form only a small subset of known stars hosting exoplanets with measured radii and masses.  In this paper, we analyze the mass--radius (M-R) relationship of planets around M dwarfs using M-R measurements for 24 exoplanets. In particular, we apply both parametric and nonparametric models and compare the two different fitting methods. We also use these methods to compare the results of the M dwarf M-R relationship with that from the \textit{Kepler} sample. Using the nonparametric method, we find that the predicted masses for the smallest and largest planets around M dwarfs are smaller than similar fits to the \textit{Kepler} data, but that the distribution of masses for 3 $R_\oplus$ planets does not substantially differ between the two datasets. With future additions to the M dwarf M-R relation from the \textit{Transiting Exoplanet Survey Satellite} and instruments like the Habitable zone Planet Finder, we will be able to characterize these differences in more detail. We release a publicly available \texttt{Python} code called \texttt{MRExo}\footnote{\url{https://github.com/shbhuk/mrexo}} which uses the nonparametric algorithm introduced by \cite{ning_predicting_2018} to fit the M-R relationship. Such a nonparametric fit does not assume an underlying power law fit to the measurements and hence can be used \replaced{on samples spanning large mass and radii ranges. Therefore by not assuming a functional form, the fit is less biased. }{to fit an M-R relationship that is less biased than a power-law.} In addition \texttt{MRExo} also offers a tool to predict mass from radius posteriors, and vice versa. 
\end{abstract}

\keywords{planets and satellites: composition}

\section{Introduction} 
\label{sec:intro}
In the Galaxy, M dwarfs are the most common type of star \citep[$\sim$ 75\%;][]{henry_solar_2006}. With the launch of the \textit{Transiting Exoplanet Survey Satellite} \citep[\textit{TESS;}][]{ricker_transiting_2014}, the hope is that we will soon discover hundreds of exoplanet candidates around them.  While the discovery of these planets is interesting in itself, the comparison between them and the \emph{Kepler} planets provides insight into the differing formation pathways of planets around M and FGK stars  \citep{lissauer_planets_2007}. For example, the pre-main-sequence lifetime of the star varies with its mass, which has an effect on the planetary migration process during its formation. Young M dwarfs are extremely active and exhibit high-intensity XUV radiation, which affects the inner planets and can potentially strip away the atmospheres of gaseous planets to leave rocky cores \citep[][show this for \textit{Kepler} planets]{owen_evaporation_2017, owen_photoevaporation_2018}. There is also empirical evidence from \textit{Kepler} \citep{dressing_occurrence_2015, gaidos_they_2016} and radial velocity (RV) surveys \citep{bonfils_harps_2013} that suggests that the mass and radius distribution of planets is not identical for M and FGK dwarfs.  These suggest that different physical processes may be at play and pose a number of questions:  Do smaller planets around M dwarf  have more rocky compositions \citep{mulders_increase_2015, mulders_stellar-mass-dependent_2015}? Is planet formation more efficient around M dwarfs \citep{dressing_occurrence_2015, ballard_kepler_2016, ballard_predicted_2018}? If so, how does planet formation impact exoplanet chemical composition?

Probabilistic mass--radius (M-R) relationships provide us with an empirical window into these questions, as they are closely related to distributions of exoplanet compositions.  They also have very practical uses.  For example,  efficient planning of \textit{TESS} follow-up RV observations requires an estimate of the planetary mass given a planetary radius. Future microlensing space missions like \textit{WFIRST} \citep{green_wide-field_2012} will produce hundreds of exoplanets with the inverse problem of having a mass but not a radius. The  M-R  relationships can be used to predict one quantity from the other. For this purpose, one needs a model for the M-R relationship that best balances the trade-off between the prediction's variance (i.e. the width of the range of possible masses for a given planet) and its bias in the predicted values (i.e. the difference between the true mass and the mean predicted mass). 

 Using transit spectroscopy due to the lower stellar brightness
and relatively large planet-to-star radius ratio, M dwarfs will
also offer potential targets for atmospheric characterization of
Earth-like habitable zone planets. It has also been shown that the spot-induced RV jitter is reduced in the near infrared (NIR) \citep{marchwinski_toward_2015}, which is where the spectral energy distributions for these stars peak. For these reasons, many more M dwarf planets will have their masses measured in the near future by instruments such as the HPF \citep{mahadevan_habitable-zone_2012},  CARMENES \citep{quirrenbach_carmenes:_2016},  NIRPS \citep{wildi_nirps:_2017}, IRD \citep{kotani_infrared_2018}, SPIRou \citep{artigau_spirou:_2014}, iSHELL \citep{cale_precise_2018}, GIANO \citep{claudi_giarpstng:_2017}, and NEID \citep{schwab_design_2016}. Here we set the stage for these future datasets by assessing the dependence of the M-R model choices on mass and radius predictions, especially as a function of stellar type.

Substantial past efforts have been put toward
studying M-R relations in recent years. A summary of this is presented in \cite{ning_predicting_2018}. Several of the widely used M-R relationships include those of \cite{ weiss_mass-radius_2014, wolfgang_probabilistic_2016, chen_probabilistic_2017}, the latter of which also introduced a publicly available \texttt{Python} package called \texttt{Forecaster}.  The M-R model underlying \texttt{Forecaster} uses a broken power law to fit the M-R relationship across a vast range of masses and radii, in recognition of potential changes in the physical mechanisms responsible for the planetary formation at different mass regimes. However, as has been discussed in \cite{ning_predicting_2018} such a restrictive parametric model can portray an incomplete picture, since we do not know the true functional form of the underlying relationship, such as whether it is a power law to begin with. Conversely, a nonparametric model offers more flexibility in the fit, which can be advantageous when the goal is to obtain predictions that best reflect the existing dataset. \cite{ning_predicting_2018} introduced a nonparametric model for the M-R relationship which uses Bernstein polynomials, a series of unnormalized Beta probability distributions.  We apply this model to  M dwarf exoplanets in preparation for future, larger datasets of small planets around small stars, which are less likely to be fully described by highly parametric models. 

While adapting the methodology of \cite{ning_predicting_2018} to a new dataset, we have developed and are now offering a publicly available \texttt{Python} package called \texttt{MREXo}, inspired by the useful community tool \texttt{Forecaster}. Not only can \texttt{MREXo} be used as a predicting and plotting tool for our M dwarf and \textit{Kepler} M-R relationships, it can also be used to fit an M-R relationship to any other dataset (but see \S \ref{sec:simulation} for a discussion about the minimum dataset size at which a nonparametric fit becomes useful). This makes it a powerful tool for exoplanet population studies and a probe for potential differences in composition across samples. This package uses the open-source tools of \texttt{Python} deployed with parallel processing for efficient computation. It also offers a fast predictive tool for the M dwarf and \textit{Kepler} sample M-R fits used in this paper, so that either mass or radius can be predicted from the other. 

The rest of the paper is structured as follows. In \S \ref{sec:data} we discuss the input dataset, and in \S \ref{sec:fitting} we discuss the parametric and nonparametric fitting process and algorithms followed. In \S \ref{sec:mrexo}, we describe the \texttt{Python} package \texttt{MRExo} that we release along with this paper. Then, in \S \ref{sec:results} we discuss the results of the fits, where we compare the parametric fit with the nonparametric and the M dwarf M-R relationship with that from the \textit{Kepler} exoplanet sample. In \S \ref{sec:simulation}, we explain the simulation performed to test the efficacy of the nonparametric method. We end with a discussion in \S \ref{sec:discussion}  and conclude in \S \ref{sec:conclusion}.

\begin{figure*}[] 
\centering
\includegraphics[width=12cm]
{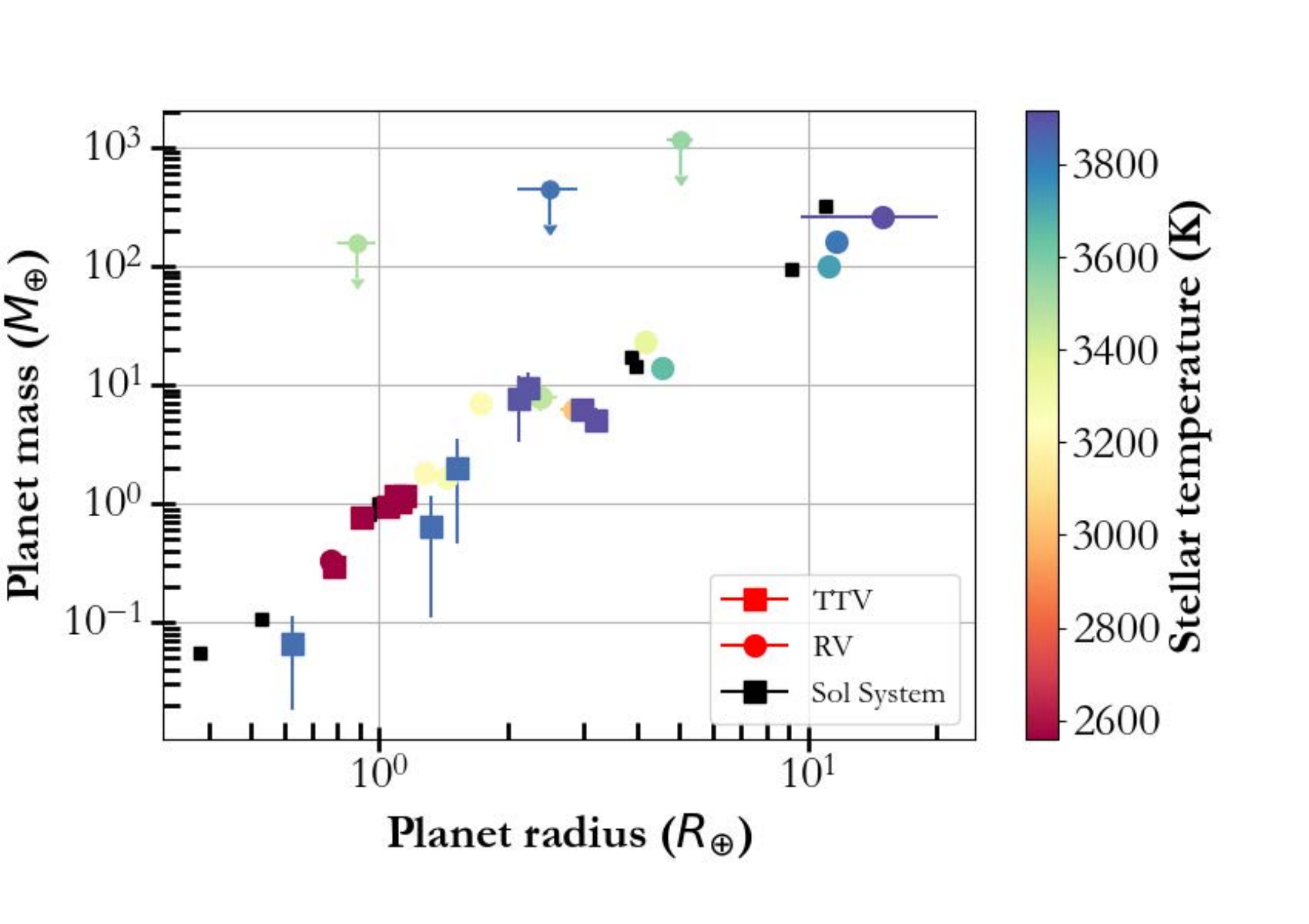}
\caption{Mass and radius of the 24 exoplanets in the M dwarf sample set, colour coded by the stellar temperature. The upper limits are shown by the arrows; these upper limits are included only in the parametric hierarchical Bayesian modelling, and not in the nonparametric fitting. The black squares represent the eight planets in our Solar system, which are shown for comparison purposes and are not used in the fitting.  See Table \ref{tab:data} for a detailed list of these planets.}
\label{fig:data}
\end{figure*}

\section{Data Set}
\label{sec:data}

Fitting an M-R relationship requires a sample set with confirmed mass and radius measurements\footnote{The nonparametric framework employed here cannot handle measurement upper limits, since the MLE method used in \citet{ning_predicting_2018} does not allow censored data; hence, planets with only upper limits are excluded from the results presented in \S \ref{sec:nonpar_fit}.  Adapting this methodology to include upper limits is an area for future work.}. The mass and radius values for the exoplanets used in this work are obtained from the NASA Exoplanet Archive, which we last accessed on 14th December 2018 \citep{akeson_nasa_2013}. \autoref{fig:data} shows the 24 M-R data points that we have used, colour-coded by host star temperature. The mean values for mass and radius, along with their respective measurement uncertainties, are shown in Table \ref{tab:data} in the Appendix. We include mass value estimates from both RV and Transit Timing Variations (TTVs). \added{The orbital periods for these planets range from about 1.5 to 33 days.}  This sample is hereafter referred to as the M dwarf dataset in this paper. We specifically chose to omit the three planets discovered by direct imaging with both radius and mass constraints, as these planets have substantially larger orbital separations than the other planets in our sample. Furthermore, the directly imaged planets have their masses and radii modelled and not directly measured.  

To limit ourselves to the M dwarfs, we restrict our host star sample to $T_{\rm{eff}}$ $<$ 4000 K.  To exclude brown dwarf companions, we restrict ourselves to planetary masses ($M_{\rm{p}}<$ 10 $M_{\rm{J}}$). We chose to use the most recent mass and radius values from the Exoplanet Archive rather than the default values. We manually verified the reported values from the Exoplanet Archive with the literature references. For the TRAPPIST-1 system, we used the \cite{grimm_nature_2018} values for the mass and radius measurements. Note that of the 24 exoplanet measurements in our sample, seven belong to the TRAPPIST-1 system and thus our M-R joint distribution is heavily influenced by the TRAPPIST-1 system in the small planet ($R$ $<$ 2 $R_{\oplus}$) regime. For purposes of comparison, we also use the \textit{Kepler} dataset from \cite{ning_predicting_2018} to compare our M dwarf results with those from a larger \textit{Kepler} sample, which consists of 127 M-R measurements and does not have any  $T_{\rm{eff}}$ restrictions on the host stars. This sample is hereafter referred to as the \textit{Kepler} dataset in this paper.

\section{Fitting the Data set}\label{sec:fitting}
To fit these data sets, we use two recently proposed methods: a parametric method proposed by \cite{wolfgang_probabilistic_2016}, and a nonparametric method by \cite{ning_predicting_2018}. The  two methods and their M-R fits are detailed and compared in subsequent sections.
We perform the following analysis. 
\begin{enumerate}
    \item Comparing the parametric and nonparametric results for the M dwarf sample.
    \item Comparing the M dwarf and \textit{Kepler} sample M-R relations using the nonparametric model.
    \item \added{Estimating the impact of the TRAPPIST-1 system on the M dwarf M-R relationship.}
    \item Performing a simulation study to show the efficacy of the nonparametric method as (a) the sample size increases, and (b) the intrinsic astrophysical dispersion in the sample increases (assuming a power law with an intrinsic dispersion).
\end{enumerate}

We compare the parametric method with the nonparametric technique because the planetary masses and radii of the M dwarf sample currently appear to resemble a power law. This may not remain the case as we accumulate more data for M dwarf planets; indeed, with over 100 mass and radius measurements, the \emph{Kepler} dataset is not currently well described by a power-law model \citep{ning_predicting_2018}.  Additionally, nonparametric methods offer less biased predictions, as long as the dataset they were fit to is representative of the underlying distribution. In contrast, nonparametric methods often perform poorly for small data sets and can easily overfit the data, while parametric methods are easier to implement and are valuable for gaining an intuitive understanding of the dataset. Acknowledging both the simplicity and easy insight from a parametric model and the improved predictive capabilities of a nonparametric model for larger datasets, we compute both and compare the outputs for the current M dwarf sample (see \S \ref{sec:results}). In addition, we run simulations to assess at what dataset size  does the nonparametric technique work well (we also test the fitting as the intrinsic dispersion in the dataset increases).

We note that the nonparametric method cannot handle upper limits in its current implementation. In the \textit{TESS} era of planet discoveries, we will soon have many exoplanet candidates with just upper limits, and so this is a clear area for future statistical development of this nonparametric approach.  For the time being, we incorporate the upper limits in the parametric approach, which can accommodate censored data like upper limits.

\subsection{Parametric method: Hierarchical Bayesian Modelling }
\label{sec:par}

The parametric method we use is heavily borrowed from the probabilistic  hierarchical Bayesian model (HBM) introduced in \cite{wolfgang_probabilistic_2016}. This model
consists of a power law to describe the mean planet mass as a function of radius plus a normal distribution around that line to describe the intrinsic astrophysical scatter in planetary masses at a specific radius.  The model we employ here is very similar to that of \citet[][Equation 2]{wolfgang_probabilistic_2016}, except that we use a lognormal distribution to describe the intrinsic scatter.  This change causes the intrinsic dispersion to be constant in log space, which \citet{chen_probabilistic_2017} showed is a better descriptor of the scatter over a larger mass range than was considered in \citet{wolfgang_probabilistic_2016}.  The parametric model used here is therefore:
\begin{equation}
\label{eq:hbm}
\log\frac{M}{M_{\oplus}} \sim \mathcal{N} \left( \mu = C + \gamma \log \frac{R}{R_{\oplus}}, \ \sigma = \sigma_M \right),
\end{equation}

\noindent where $C$ is the normalization constant of the mean power law (once it is converted to the linear mass scale), the $\gamma$ is the power-law index, and the $\sigma_M$ is the intrinsic scatter in terms of log($M$).  The $\sim$ symbol implies that the masses are drawn from a probabilistic distribution (as opposed to an $=$ sign for a deterministic model); this distribution is what models the astrophysical scatter in planetary masses.  Here it is parameterized as a Gaussian centered on a line, which when converted to linear scale is a lognormal distribution centered on a power law.  Evaluating this model within a hierarchical framework allows the upper limits in our dataset to be incorporated into the inference of $C$, $\gamma$, and $\sigma_M$, along with an arbitrary measurement uncertainty for individual data.

\subsection{The Nonparametric Method -- Bernstein polynomials  } \label{sec:nonpar}
As explained in the Introduction, we also employ the nonparametric model introduced by \cite{ning_predicting_2018} to fit  M-R relationships, with an eye toward future M dwarf dataset sizes that will likely benefit from a more flexible approach. The nonparametric method fits the joint M-R distribution, and hence can treat either mass or radius as the independent variable, and can be used to predict one from the other. 
This method uses a sequence of Bernstein polynomials as the basis functions to fit a nonparametric M-R relationship; when normalized, these Bernstein polynomials are identical to beta probability distributions. Hence, the model is equivalent to a mixture of unnormalized Beta probability distributions. We translate the code presented in \citet{ning_predicting_2018} from \texttt{R} to \texttt{Python} and release it in a publicly available package called \texttt{MRExo}; this code is further discussed in \S \ref{sec:mrexo}.

The nonparametric approach uses these Bernstein polynomials as the basis functions to fit the joint distribution of masses and radii \textit{f(m,r)}. This is detailed in \S 2 of \cite{ning_predicting_2018}:
\begin{align}
\label{eq:nonpar_joint}
& f(m,r|w,d,d') 
\nonumber \\
& \quad = \sum_{k=1}^{d} \sum_{l=1}^{d'} w_{kl} \frac{\beta_{kd}\left(\frac{m - \underline{M}}{\overline{M} - \underline{M}} \right)}{\overline{M} - \underline{M}}  \frac{\beta_{ld'}\left(\frac{r - \underline{R}}{\overline{R} - \underline{R}} \right)}{\overline{R} - \underline{R}},
\end{align}

\begin{table*}[]
\begin{align}
& f(m,r | \textbf{\textit{w}}, d, d', M^{obs}, R^{obs} , \mathbf{\sigma}_M^{obs}, \mathbf{\sigma}_R^{obs}) 
\nonumber\\
&\quad = \prod\limits_{i=1}^{n} f(M_i^{obs}|m,\sigma_{M_{i}}^{obs}) f(R_i^{obs}|r,\sigma_{R_{i}}^{obs}) \times f(m,r|w,d,d') \\ 
&\quad = \prod\limits_{i=1}^{n} \sum\limits_{k=1}^{d} \sum\limits_{l=1}^{d'} w_{kl} \frac{\beta_{kd}\left(\frac{m - \underline{M}}{\overline{M} - \underline{M}} \right)}{\overline{M} - \underline{M}} \mathcal{N}\left(\frac{M_i^{obs} - m}{\sigma^{obs}_{M_i}} \right) \times \frac{\beta_{ld'}\left(\frac{r - \underline{R}}{\overline{R} - \underline{R}} \right)}{\overline{R} - \underline{R}} \mathcal{N}\left(\frac{R_i^{obs} - r}{\sigma^{obs}_{R_i}} \right),
\label{eq:joint_w_uncertain}
\end{align}
\end{table*}

\noindent where $w_{kl}$ is the $kl$-th element of the matrix \textit{w},  which is a set of weights corresponding to individual Bernstein polynomials. Each weight is positive, and the sum of individual weights equals unity. The weights describe how much each term in the series contributes to the joint distribution. The degrees for the polynomials in the mass and radius dimensions are represented by \textit{d} and  \textit{$d'$}, respectively. Here \textit{m} and \textit{r} depict the continuous variables which represent mass and radii. $\underline{M}$, $\overline{M}$, $\underline{R}$ and $\overline{R}$ represent the lower and upper bounds in mass and radius, respectively. To fit for the weights ($w_{kl}$), we use maximum-likelihood estimation (MLE).  The likelihood (\autoref{eq:joint_w_uncertain}) includes measurement uncertainties as normal distributions; hence,  \autoref{eq:joint_w_uncertain} modifies \autoref{eq:nonpar_joint} to introduce a convolution of the normal and Bernstein polynomials.  The measured values for mass and radii are assumed to be drawn from a normal distribution centered on the true value, with a standard deviation equal to the measurement uncertainty. This produces the joint distribution shown in \autoref{eq:joint_w_uncertain}, \noindent where $M_i^{obs}$, $\sigma_{M_{i}}^{obs}$ are the mass observations and their uncertainties, and likewise $R_i^{obs}$, $\sigma_{R_{i}}^{obs}$ for radius. Here $\mathcal{N}$ denotes a normal distribution. After optimization for the weights using the MLE method via the \texttt{Python} package \texttt{SciPy} \citep{oliphant_python_2007}, the joint distribution for the M dwarf dataset can be plotted (see \autoref{fig:results} b).  

A key consideration in fitting this nonparametric model to a dataset is identifying the optimum degree for the Bernstein polynomial series. We use the cross-validation method as explained in \citet{ning_predicting_2018} to find that the optimum values for \textit{d} and  \textit{$d'$} are both 17. In total, there are $\textit{d} \times$ \textit{$d'$} weights that serve as our `parameters' to fit for. While $17^2=289$ parameters may seem excessive for a dataset size of 24 planets, in practice, there are only five non-zero weights in the series (see \autoref{fig:m_weights} for a pictorial representation of this).  This highlights one of the advantages of using Bernstein polynomials as our basis functions: estimating the weights in this series is self-regularizing, meaning that the smallest number of nonzero coefficients is automatically found.  Additionally, using Bernstein polynomials efficiently reduces the number of nonzero free parameters; if one instead used a mixture of Gaussians to fit the joint distribution, that fit would require at least three times as many free parameters (amplitude, mean, and standard deviation per component, rather than just the polynomial coefficients).  

After finding the weights via MLE, we repeat the process using the bootstrap method, which \replaced{estimates the uncertainty in this fit.}{helps to account for the variation of the parameters in the model.} We do this by resampling the dataset with replacement and running the fitting routine again for each bootstrap.  In regions without data points, the Bernstein polynomials revert to the overall mean of the distribution.

We found that the Bernstein polynomials may behave counter-intuitively at the boundaries of the joint distribution  (see Appendix A in \cite{ning_predicting_2018}). To address this issue, we fix the first and last row and column of Bernstein polynomials to have zero weights ($w_{0j}, w_{dj}, w_{i0}, w_{id'}$), and as such, they do not contribute to the joint distribution. This was not done by \cite{ning_predicting_2018} since their \textit{Kepler} sample had enough samples near the boundaries to constrain the fit.

\section{MRExo} \label{sec:mrexo}
In this section, we shall discuss an important contribution made in this paper. We translate \cite{ning_predicting_2018}'s \texttt{R} code \footnote{\noindent\url{https://github.com/Bo-Ning/Predicting-exoplanet-mass-and-radius-relationship}.} into a publicly available \texttt{Python} package called \texttt{MRExo}\footnote{\url{https://github.com/shbhuk/mrexo}}. This can be easily installed using \texttt{PyPI} and has extensive documentation and tutorials to make it easy to use.

The \texttt{MRExo} package offers tools for fitting the M-R relationship to a given data set. In this package, we use a cross-validation technique to optimize for the number of degrees.  We then fit the joint distribution ($\S$ \ref{sec:nonpar}) to the sample set; this can then be marginalized to obtain the conditional distribution, which we can use to predict one variable from the other. We bootstrap our fitting procedure to estimate the uncertainties in the mean M-R relation.  Further, \texttt{MRExo} is equipped with dedicated and easy-to-use functions to plot the best-fit conditional  M-R relationships, as well as the joint M-R distribution\footnote{We caution the user not to overinterpret the joint distribution; in the first version of this software we have made no attempt to correct for detection and selection bias, which is needed before the joint distribution can be interpreted as an M-R occurrence rate.  We show the joint distribution in \autoref{fig:results} solely to illustrate how the behavior of the conditionals relates to the joint.}.  Crucially, \texttt{MRExo} also predicts mass from radius, and radius from mass.  For example, in the case of planets discovered using the transit method, the feasibility of an RV follow-up campaign can be evaluated by predicting the estimated mass and its confidence intervals given the measured radius and its uncertainty. Another feature of this package is that it can accommodate radius (or mass) posterior samples from separate analyses, which are then used to compute the posterior predictive distribution for mass . Along with the \texttt{MRExo} installation, the results from the M dwarf sample dataset from this paper and the \textit{Kepler} exoplanet sample from \cite{ning_predicting_2018} are included.

The degree of the Bernstein polynomials ($d$) approximately scales with the sample size ($N$). Since the number of weights goes as $d^2$, the computation time involved in the fitting a new M-R can soon start to become prohibitive. Therefore, we also parallelize the fitting procedure and the bootstrapping algorithm.  As an example, the M dwarf sample ($N=24$; $d=17$) took about 2 minutes to perform cross-validation, fit a relationship, and do 100 bootstraps on a cluster node with 24 cores and 2.2 GHz processors. The \textit{Kepler} sample ($N=127$; $d=55$) took about 36 hr for the cross-validation, fitting, and 48 bootstraps. We realize that the fitting computation time will start to become prohibitive as the sample size increases $\gtrsim 200$; therefore, we plan to optimize the code further by benchmarking, optimizing floating point operations, and correcting the precision requirements in the integration step. However, this time-intensive step of cross-validation and fitting is only necessary if the user needs to fit their own relationships. To run the prediction routine on the preexisting M dwarf or \textit{Kepler} samples is fairly quick and takes less than a second for a prediction. In order to do a large number of predictions as part of a larger pipeline or simulation, the user can also generate a lookup table that makes the calculations even faster (the function to generate and use the lookup table is provided with the package).

\section{Results}\label{sec:results}

\subsection{Parametric fit results}\label{sec:par_fit}

As discussed in \S \ref{sec:fitting}, there is a trade-off between the flexibility and lower bias of nonparametric models and the interpretability and lower predictive variance of parametric models.  At present, the M dwarf dataset only consists of 24 planets, and their masses and radii seem to be well approximated by a power law.  As such, we fit the parametric model described by \autoref{eq:hbm} to the same M dwarf data set, to serve as a basis for comparison to the nonparametric results.  Currently, the fit looks to be fairly reasonable by eye (see \autoref{fig:results}e), but this is not guaranteed to be the case as more measurements are obtained (indeed, it is clear from the fit to the \textit{Kepler} data, which consist of $>100$ planets, that a single power law is not a sufficient model for that data; see \autoref{fig:results}f).

The best-fit parameter values for the model described in \autoref{eq:hbm} are $C = -0.130^{+0.081}_{-0.055}, \gamma = 2.14^{+0.11}_{-0.16}$, and $\sigma_M = 0.184^{+0.077}_{-0.021}$; these values were found by identifying the mode of the joint three-dimensional hyperparameter posterior, and the error bars represent the marginal central 68\% credible intervals.   This power-law slope is steeper than that found by \citet{wolfgang_probabilistic_2016} for nearly all of the datasets they consider (most lie within $1.3 < \gamma < 1.8$).  There are two possible explanations for this.  First, the dataset used by \citet{wolfgang_probabilistic_2016} --- and, in fact, nearly all previous M-R results except \citet{neil_host_2018} --- is dominated by FGK dwarf planet hosts.  This result could therefore be driven by intrinsic differences between the planets that form around Sun-like stars and those that form around M dwarfs.  However, \citet{wolfgang_probabilistic_2016} also use only a limited radius range for their dataset ($R < 4$ $R_\oplus$ or $R < 8$ $R_\oplus$), as their focus was on super-Earths.  As shown by \citet{ning_predicting_2018}, the slope for $1$ $R_\oplus < R < 4$ $R_\oplus$ is shallower than that for $4$  $R_\oplus < R < 10$ $R_\oplus$, and so one would expect that a single power-law fit to the entire radius range would result in a steeper slope than that fit just to $R < 4$ $R_\oplus$.  This is corroborated by the fact that the slopes fit to the $R < 8$ $R_\oplus$ radius range are steeper than those fit to the $R < 4$ $R_\oplus$ range.  Additionally, \citet{chen_probabilistic_2017} found that
the segment of their broken power-law relation that most closely matches radius range of our M dwarf dataset ($1$ $R_\oplus \lesssim R \lesssim 11$ $R_\oplus$) has a power-law slope of $\frac{1}{0.59} = 1.7$.  We note that the fact that there is a noticeable difference between the slopes fit to different radius ranges is evidence that a more flexible relation, such as the nonparametric one developed in \citet{ning_predicting_2018} is needed. \added{A broken power-law approach, such as the one implemented by \citet{chen_probabilistic_2017} could be employed; however, a nonparametric method provides for a smooth transition between adjacent radius ranges, something the broken power law does not.}

Another notable difference between \replaced{these two relations}{the parametric fit for the M dwarf planets and the FGK planet sample from \citet{wolfgang_probabilistic_2016}} is the predicted mass for a 1 $R_\oplus$ planet.  This information is contained in $C$, the mean power-law normalization constant.  In linear units, the best-fit $C$ for the M dwarf dataset is $10^{-0.13} = 0.74$ $M_\oplus$.  This is significantly lower than the power-law constants found in \citet{wolfgang_probabilistic_2016}: it is inconsistent with the $R < 8$ $R_\oplus$ fit ($C = 1.5$ $M_\oplus$) at 4$\sigma$, and with the $R < 4$ $R_\oplus$ fit ($C = 2.7$ $M_\oplus$) at 7$\sigma$.  At least two effects are contributing to this difference.  First, very few planets around FGK dwarfs with $R < 1$ $R_\oplus$ have had their masses measured and constrained to be $M < 1.5$ $M_\oplus$.  Because of the publication bias quantified in \citet{burt_simTESSfollow_2018}, wherein planetary masses are required to reach a certain significance threshold to be published, small, low-mass planets are more likely to be left out of the FGK dwarf datasets.  This causes the relation to be fit to the more massive small planets that do appear in the literature, which in turn causes $C$ to be high.  Conversely, the M dwarf dataset is strongly affected by the presence of the TRAPPIST-1 planets \citep{gillon_trappist_2017}, whose masses are reported by \citet{grimm_nature_2018} to be both small and quite precise.  This one analysis of a single system dominates the M dwarf dataset at $0.7$ $R_\oplus$ $< R < 1.2$ $R_\oplus$ and therefore the best-fit value of $C$ --- a caveat that all potential users of these M dwarf relations should keep in mind. \added{We study the impact of the TRAPPIST-1 system on our M-R fits in detail in $\S$ \ref{sec:wo_Trappist}.}

To determine whether the differences highlighted above are due to the differences in the host star population, models, or considered radius ranges, we also fit this simple parametric model to the \textit{Kepler} dataset from \citet{ning_predicting_2018}.  While we do not expect this model to be a good fit to the data, it offers an apples-to-apples comparison to the above M dwarf parametric results, and its shortcomings highlight the advantages of more flexible nonparametric models.  The best-fit parameter values for the Kepler dataset are $C = -0.0250^{+0.093}_{-0.111}, \gamma = 2.13^{+0.11}_{-0.14}$, and $\sigma_M = 0.457^{+0.026}_{-0.044}$.  Therefore, there is no statistically significant difference between the M dwarf planetary mean power law  and the one that would be fit to the planets around FGK dwarfs.  More specifically, the two slopes are nearly identical; while the \textit{Kepler} normalization constant is larger than the M dwarf constant, the difference is at the $1\sigma$ level.  This comparison is consistent with the result of \citet{neil_host_2018}, which used a sample size of six M dwarf planets.  On the other hand, there is a statistically significant difference in the intrinsic scatter for the M dwarf and \textit{Kepler} datasets, with the \textit{Kepler} dataset having more variation in planet mass at a given radius.  However, this may be at least partially affected by the differences in the sample sizes ($N=24$ vs.\ $N=127$) and the fact that it takes more time to get significant mass measurements for planets on the low mass side of the intrinsic scatter distribution.  As we obtain more masses for transiting M dwarf planets with instruments like HPF, we will be able to test whether the difference in the intrinsic scatter remains statistically significant.

All told, the comparison between the M dwarf and \textit{Kepler} parametric model fits does not yield very much insight into the differences between the planetary populations.  Because the parametric model is not able to capture detailed features of the M-R relation, it may hide some important differences between the two datasets once the M dwarf dataset becomes large enough to warrant detailed analyses of these features.  Looking ahead to the M-R dataset we will have by the end of the \textit{TESS} mission,  we apply the nonparametric model to the current M dwarf dataset and perform the comparison between it and the nonparametric fits to the \textit{Kepler} dataset to serve as a basis for comparison to this future work (see \S \ref{sec:compare_m_kepler}).

\subsection{Nonparametric fit}\label{sec:nonpar_fit}

\begin{figure}[] 
\centering
\includegraphics[width=\columnwidth]
{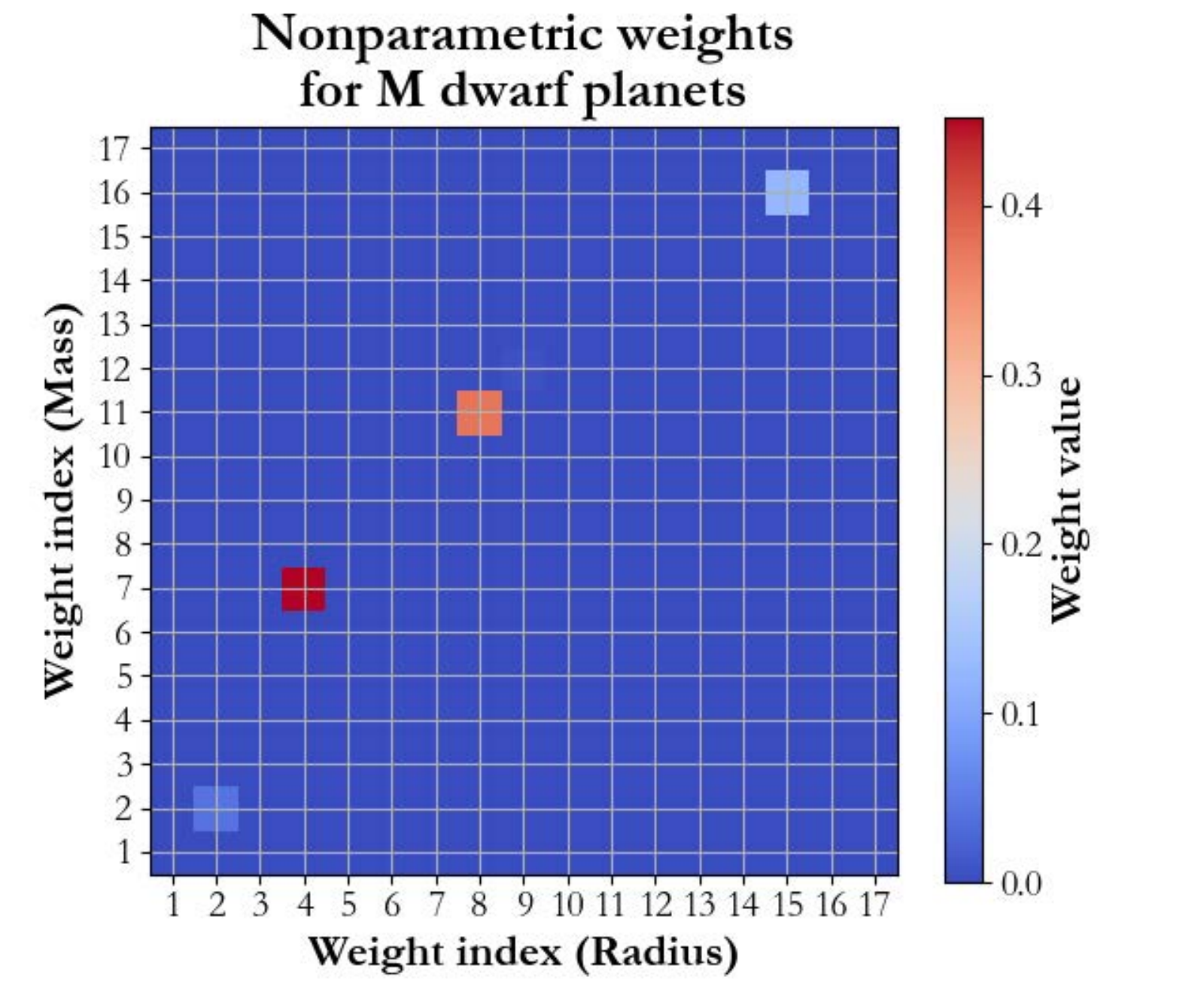}
\caption{Optimal weights chosen by cross-validation for the M dwarf sample. The cross-validation procedure selected 17 degrees as the optimum for both mass and radius which gives 17$^2 = 289$ weights, of which only five (four are visible, the fifth one at 9,12 is difficult to see in this stretch) are nonzero.} \label{fig:m_weights}
\end{figure}

\begin{figure*}[!t] 
\centering
\includegraphics[width=18cm, angle = 90]
{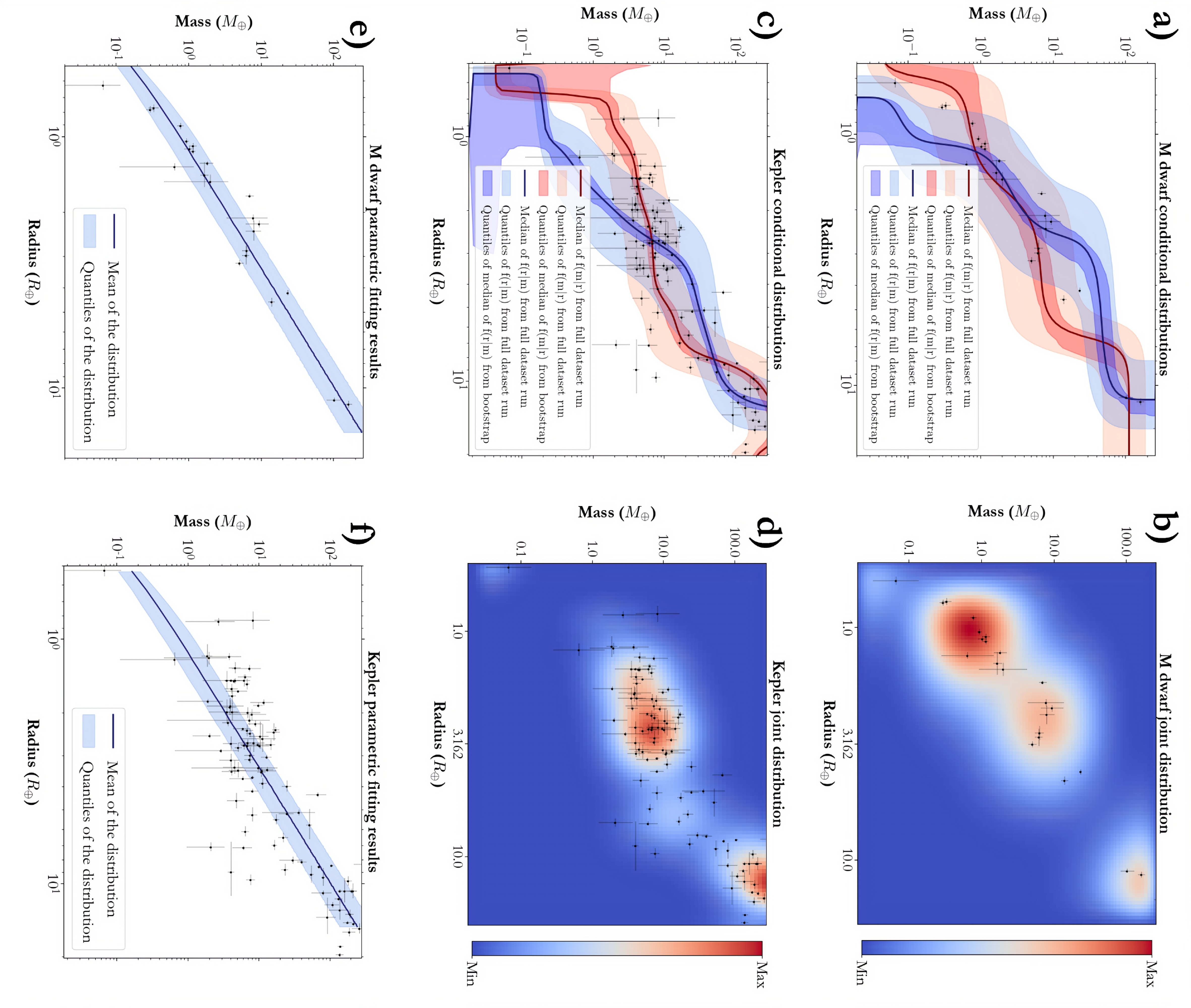}
\caption{\textbf{Panels (a), (b), (c), (d):} nonparametric fitting results for the M dwarf dataset (see  \autoref{fig:data}) and \textit{Kepler} sample. \textbf{Panels (e) and (f)):} parametric fitting results for the M dwarf and \textit{Kepler} sample. \textbf{(a)} The conditional distributions $f(m|r)$ and $f(r|m)$ are shown in red and blue, respectively. The dark line represents the mean of the conditional distribution that was obtained from the full dataset run (no bootstrapping); this is the most likely mass at a given radius (or radius at a given mass).  The lightest shaded region represents the 16 - 84 $\%$ quantiles of the conditional distribution; this illustrates the width of the predicted mass (or radius) distribution at that radius (or mass). The darker shaded region is the result of the bootstrapping procedure and shows the 16 $\%$ and 84 $\%$  quantiles of the median of the distribution; this represents the uncertainty in the solid line. \added{The increased uncertainty in the bootstrap regions, especially near the boundaries, is because of the sparseness of data near the boundaries. This leaves the relation unconstrained when the dataset is resampled with replacement.} \textbf{(b)} Joint probability distribution $f(m,r)$ where the background colour represents the highest probability. \textbf{(c),(d)} The same conditional and joint distribution plot for the \textit{Kepler} sample. \textbf{(e),(f)} Posterior predictive distribution from the parametric fitting of the M dwarf and \textit{Kepler} samples using HBM. } \label{fig:results}
\end{figure*}

We fit the M dwarf exoplanet dataset shown in \autoref{fig:data} using the nonparametric approach described in \S \ref{sec:nonpar}; the results of this fit are displayed in \autoref{fig:results} (a) and (b). Using the cross-validation method the optimum number of degrees selected are both equal to 17, giving 17$^2 = 289$ total weights. We note that our algorithm automatically forces most of the weights to zero to prevent overfitting, especially for small datasets. For this dataset, only five weights are nonzero (see \autoref{fig:m_weights}).

We also plot the conditional distributions $f(m|r)$ and $f(r|m)$, where the conditional distribution $f(m|r)$ is the ratio of the joint distribution (\autoref{eq:nonpar_joint}) to the marginal distribution of the radius $f(r)$; likewise for $f(r|m)$(see Equations 1 and 2 of \citet{ning_predicting_2018}). The conditional distribution spans regions of M-R space where there is no data. This is due to the symmetrical nature of the Bernstein polynomials and indicates that the current dataset is not yet large enough to be effectively described by a nonparametric method; this is further discussed in \S \ref{sec:simulation}.

From \autoref{fig:results}(b), we note that the red conditional distribution--$f(m|r)$  is not the same as the blue conditional distribution--$f(r|m)$. This is because when the joint distribution (see \autoref{fig:results} (b) and (d)) is marginalized to obtain the conditional, the distribution behaves differently for each axis. The two conditional distributions would be the same for a dataset with very little error  and symmetric, equal scatter along the entire M-R relation (see \S \ref{sec:simulation} for an illustration of this).  The \emph{Kepler} dataset effectively illustrates how localized areas of large scatter in one dimension can drive differences between $f(m|r)$ and $f(r|m)$.  In particular, at log($m$) $\sim 0.5$, there is large scatter in radius, which is fit by the mean M-R relation in $f(m|r)$ but is represented by the distribution (the width of the light shaded region) around that mean relation in $f(r|m)$.  While both conditionals were computed from the same joint distribution (\autoref{fig:results}d), asymmetries such as this in the joint distribution can produce conditionals that qualitatively look very different.  We also note that truncating the probability distributions for integration (that is, using finite bounds to compute $f(r)$ and $f(m)$ for the conditionals) also contributes to some differences between the two conditional distributions.  This can be further seen in the simulation in  \autoref{fig:sim_disp}, where the disparity between the two conditional distributions increases as the dispersion from the power law increases. Therefore, to predict one quantity from the other it is imperative to use the right conditional distribution.

A common concern about  nonparametric methods is their ready ability to overfit data.  The cross-validation method we adopt to choose $d$ and $d'$ is designed to minimize overfitting while maximizing the predictive accuracy of the M-R relation.  Both underfitting and overfitting produce high predictive error (that is, the predicted value is far from the true value); because the cross-validation method identifies the degree that minimizes the predictive error, it finds the optimum $d$ that balances the trade-off between the two.  That said, we acknowledge that by eye, there are wiggles in the M dwarf M-R relation that do not appear to be supported by the dataset.  This is likely due to the sparsity of the dataset in certain radius and mass ranges; where there is no data, the relation tends toward the mean of the closest nonzero term.  This is why we perform the simulation study in \S \ref{sec:simulation}, to find at what size dataset does our nonparametric model effectively describe the conditionals.

\subsection{Comparing Nonparametric vs.\ Parametric fitting}\label{sec:com_p_np}

\begin{figure*}[!t] 
\centering
\includegraphics[width=\textwidth]
{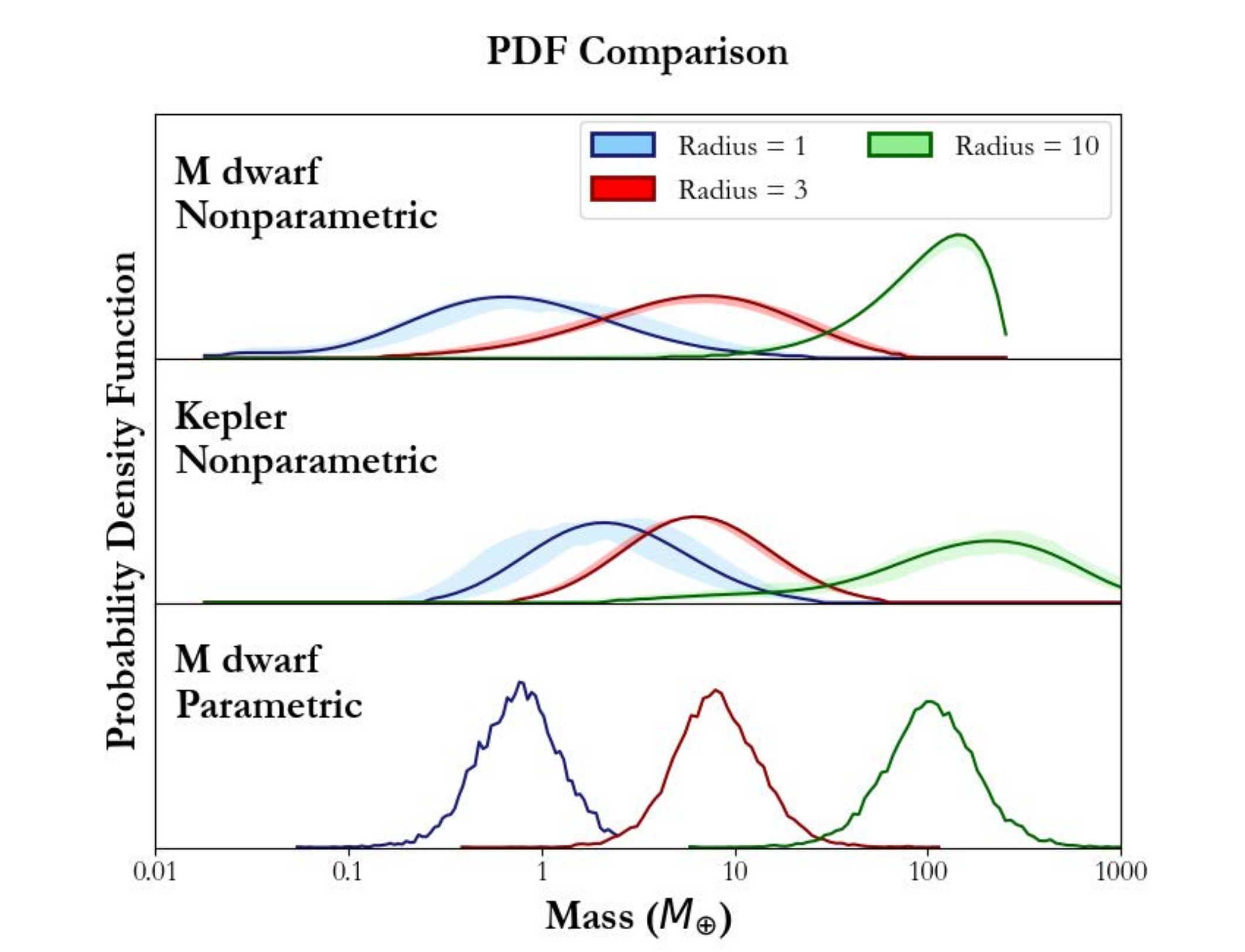}
\caption{The conditional probability density functions (CPDFs) of mass given radius for planets at 1, 3, and 10 $R_{\oplus}$; these show the predicted mass distributions at these radii from the nonparametric fit to the M dwarf sample (top), the nonparametric fit to the \textit{Kepler} sample (middle), and  the parametric fit to the M dwarf sample (bottom). The means of both M dwarf fits are at lower mass for 1 and 10 $R_{\oplus}$ planets than for the \emph{Kepler} fit, indicating that the smallest and largest planets are less massive around M dwarfs than around Sun-like stars.  Interestingly, the means of the predictive masses for 3 $R_{\oplus}$ planets are similar.  } \label{fig:pdf_compare}
\end{figure*}

We plot the conditional probability density functions (CPDFs) of planetary masses for planets with $R = 1,$ $3$, and $10$ $R_{\oplus}$ in \autoref{fig:pdf_compare}.  These CPDFs show the distribution of planetary masses that would be predicted for planets at those radii.  Assessing the differences visually, the parametric and nonparametric M dwarf fits have the same median but different spreads. This difference is expected due to the bias vs.\ variance trade-off in parametric vs.\ nonparametric modeling: if the parametric model is a poor description of the data, the nonparametric fit will be less biased but have a higher variance than the parametric estimator. Because the medians of the CPDFs are similar between the nonparametric and parametric models, it appears that, with this current dataset, a power law is a decent fit to the data.  That said, we emphasize that this will likely not remain the case as the M dwarf dataset grows, as we have seen for the \emph{Kepler} dataset.

To quantify the differences between the mass predictions produced by these two model choices, we perform the two-sample Kolmogorov-Smirnov \citep[K-S; ][]{kolmogorov_sulla_1933, smirnov_table_1948} and Anderson-Darling \citep[AD; ][]{anderson_asymptotic_1952, scholz_ADksamples_1987} tests on samples drawn from these CPDFs (see \autoref{tab:cdf_compare_p_np}).  These statistical tests assess whether we can reject the null hypothesis that two datasets were drawn from the same distribution, with the K-S test being more sensitive to the location of the median and the AD tests more sensitive to differences of the tails. 

Importantly, the reported significance of these tests depends on the size of the datasets, with small differences between the samples becoming more significant as the datasets grow.  Since we have the functional form of these CPDFs from our fitting procedure, we have a choice in how many samples we can draw from them, and therefore how significant we report them to be\footnote{For every comparison in Tables \ref{tab:cdf_compare_p_np} and \ref{tab:cdf_compare_m_kepler}, we use 100 samples from each CPDF.  We also note that the assignment of a $p$-value to a given statistic depends on a theoretical statistical distribution for the variation in that statistic given randomly drawn datasets from the same distribution.  To test this theoretically assigned $p$-value, we generated reference distributions of K-S and AD statistics based on 20,000 randomly drawn datasets (each of which also contains $N=100$ planets) from the CPDFs, compared against themselves for a total of 10,000 comparisons (and 10,000 values for the K-S and AD statistics).  We then identified at what quantile is the median of the nonparametric vs.\ parametric K-S statistic distribution with respect to the distribution of KS statistics produced by comparing the M dwarf nonparametric CPDFs against themselves.  These values are what is reported in Table \ref{tab:cdf_compare_p_np} (and an analogous comparison but between the \textit{Kepler} and M dwarf nonparametric CPDFs with respect to the self-compared nonparametric M dwarf CPDFs is reported in Table \ref{tab:cdf_compare_m_kepler}).  It turns out that this more careful approach is in good agreement with the theoretical p-values provided by the R package kSamples, built from the work of \citet{scholz_ADksamples_1987}.  If the median of the nonparametric vs.\ parametric K-S statistic distribution fell completely outside that of the nonparametric self-compared KS distribution, the \textit{p}-value is reported as $< 10^{-4}$, as we only generate 10,000 pairs of datasets for these K-S and AD statistic distributions.}.  As such, the results in Tables \ref{tab:cdf_compare_p_np} and \ref{tab:cdf_compare_m_kepler} should be interpreted on a comparative basis, not on an absolute basis: rather than concluding that the CPDFs for 1 $R_\oplus$ planets are different on a statistically significant level while the CPDFs for 10 $R_\oplus$ are not, the correct interpretation is that it will take a smaller number of mass measurements of planets at 1 $R_\oplus$ to distinguish between the parametric and nonparametric models than it will for planets at 10 $R_\oplus$ (see Table \ref{tab:cdf_compare_p_np}).  Along the same vein, an even smaller number of mass measurements would be needed to distinguish between the parametric and nonparametric models for 3 $R_\oplus$ planets.  Because we only have 24 planets in our M dwarf dataset in total, let alone at any single radius, these model comparisons will need to be reassessed in the future with more data.

\begin{deluxetable}{l|c|c|c|c}
\tabletypesize{\footnotesize}
\tablecolumns{5} 
\tablecaption{\label{tab:cdf_compare_p_np}Two-sample Comparison Tests between the Parametric and Nonparametric fits to the M dwarf sample.  Note that these are relative $p$-values based on generated datasets of 100 planets, not absolute $p$-values (see \S \ref{sec:com_p_np} for discussion).} 
\tablehead{\colhead {Radius} & \multicolumn{2}{c}{Kolmogorov-Smirnov}  & \multicolumn{2}{c}{Anderson-Darling} \\
               \colhead {$R_{\oplus}$}    & \colhead {Statistic}           & \colhead {$p$-value} & \colhead {Statistic} & \colhead {$p$-value} }
    \startdata
1              & 0.26                & 1.3 $\times$ 10$^{-3}$         & 8.30           & $<$ 10$^{-4}$               \\
3              & 0.30                & 2 $\times$ 10$^{-4}$            & 8.65           & $<$ 10$^{-4}$             \\
10             & 0.15                & 0.1511           & 1.87           & 0.1046          \\
    \enddata
\end{deluxetable}

\subsection{Comparing M dwarf M-R vs.\ \textit{Kepler} M-R}\label{sec:compare_m_kepler}
We also seek to compare the M-R relationship from a \textit{Kepler} exoplanet sample to an M dwarf exoplanet sample. As discussed in  \S \ref{sec:intro}, there is empirical evidence that the planetary radius and mass distributions differ between M and FGK dwarfs. This is further illustrated empirically by our fit to the M dwarf and \textit{Kepler} sample (\autoref{fig:results}): there are visual differences between the joint and conditional distributions for the M dwarfs and \textit{Kepler}. Additionally, \autoref{fig:pdf_compare} shows that the median value of the CPDFs of \textit{Kepler} and M dwarf masses do not coincide for the smallest and largest planets; this is true whether the nonparametric M dwarf relation or the parametric M dwarf relation is used. Therefore, if the \textit{Kepler} M-R relationship was used to predict the mass of a transiting M dwarf exoplanet like TRAPPIST-1 b, it would produce mass predictions that were too large, on average. Conversely, if used for a nontransiting planet like Proxima-b to predict its radius from its mass, the prediction would be erroneous.  This illustrates that the conditional density functions are different for the two samples. We also perform the K-S and AD tests to quantify differences between the M dwarf and \textit{Kepler} CPDFs (see \S \ref{sec:com_p_np} for discussion about these tests).  We find that it will take $<<100$ planets to distinguish between the M dwarf and \textit{Kepler} mass predictions at 1 and 10 $R_{\oplus}$, while even a dataset of 100 3 $R_{\oplus}$ planets does not illuminate any statistically significant difference between the mass predictions. Therefore, the M dwarf M-R relation could be most different from the FGK dwarf M-R relation at the smallest and the largest planet sizes. 

\begin{deluxetable}{l|c|c|c|c}
\tabletypesize{\footnotesize}
\tablecolumns{5} 
\tablecaption{\label{tab:cdf_compare_m_kepler}2-sample comparison tests between the nonparametric fit to the M dwarf and \textit{Kepler} samples. Note that these are relative $p$-values based on generated datasets of 100 planets, not absolute $p$-values (see \S \ref{sec:com_p_np} for discussion).} 
\tablehead{\colhead {Radius} & \multicolumn{2}{c}{Kolmogorov-Smirnov}  & \multicolumn{2}{c}{Anderson-Darling} \\
               \colhead {$R_{\oplus}$}    & \colhead {Statistic}           & \colhead {$p$-value} & \colhead {Statistic} & \colhead {$p$-value} }
    \startdata
1              & 0.47                & $<$ 10$^{-4}$         & 24.2          & $<$ 10$^{-4}$              \\
3              & 0.16                & 0.1057             & 2.03          & 0.0824               \\
10             & 0.42                & $<$ 10$^{-4}$          & 14.5           & $<$ 10$^{-4}$          \\ 
    \enddata
\end{deluxetable} 

\begin{figure*}[!t] 
\centering
\includegraphics[width=0.7\textwidth, angle = 90]
{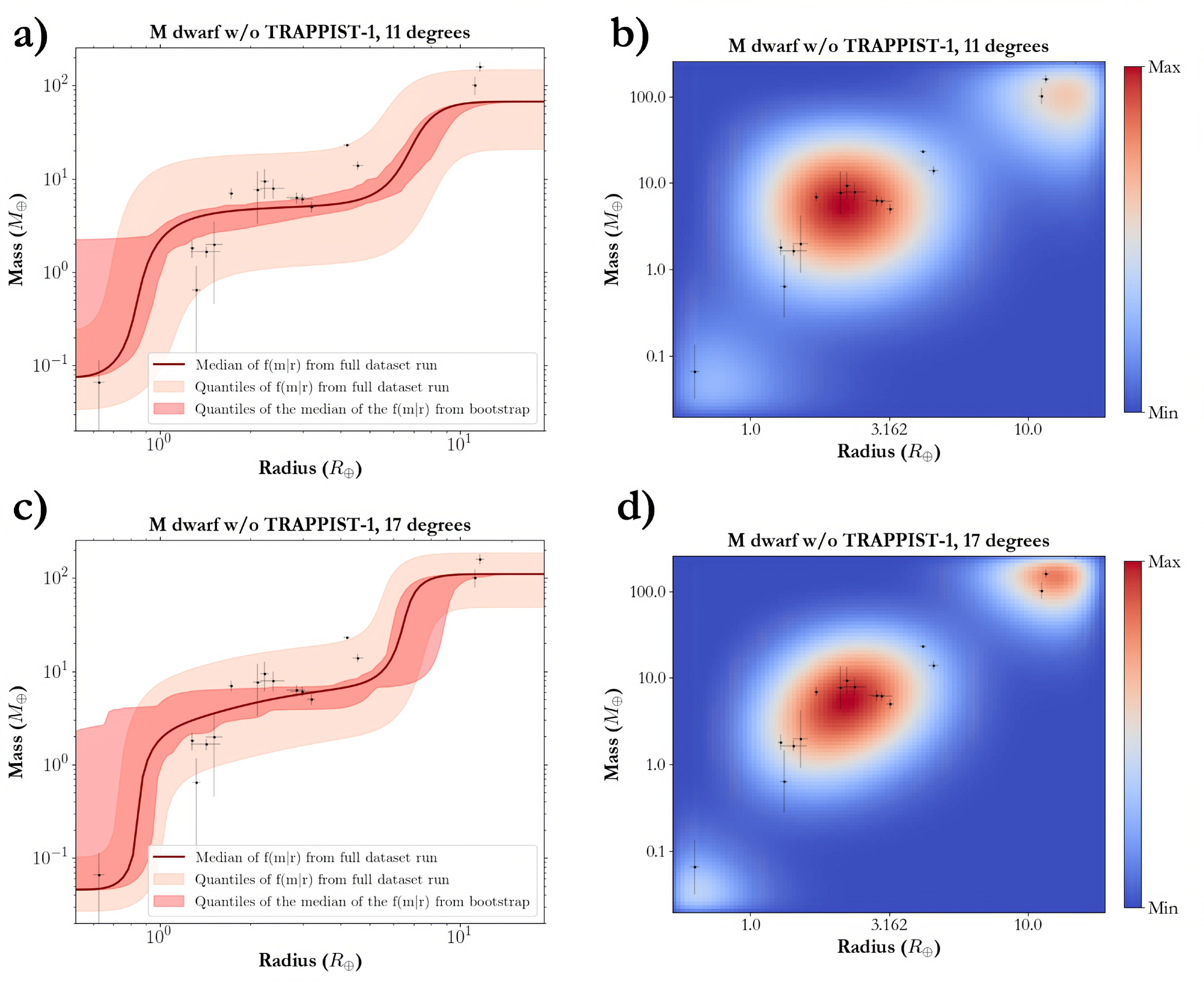}
\caption{Similar to \autoref{fig:results}, showing the conditional $f(m|r)$ ((a) and (c)) and joint ((b) and (d)) distribution for the M dwarf sample with and without the TRAPPIST-1 planets. \textbf{Panels (a) and (b)} show the case without the TRAPPIST-1 system with the M-R relationship fit using 11 degrees, as optimized by the cross-validation method. \textbf{Panels (c) and (d)} have the same for 17 degrees, which is the number of degrees the original M dwarf M-R relationship was fit with. These should be compared with the fit result plots for the M dwarf sample in \autoref{fig:results}.} \label{fig:Trappist_impact}
\end{figure*}

\added{\subsection{Estimating the impact of the TRAPPIST-1 system on the M dwarf M-R relationship}\label{sec:wo_Trappist}
Considering that our sample of M dwarf planets with masses and radii is limited to 24 planets, of which seven belong to the TRAPPIST-1 system, we explore the influence of the system on the M-R relationship in the 1 $R_{\oplus}$ regime. To do this, we eliminate the 7 TRAPPIST-1 planets from our 24 planet sample. We are left with 17 planets which we then fit in two ways- 
\begin{enumerate}
    \item We fit the sample using 17 degrees, the same number of degrees
used to fit the full 24 planet M dwarf planet sample (see $\S$ \ref{sec:nonpar_fit}).
    \item We rerun our cross-validation algorithm on this reduced sample to estimate the optimum number of degrees to use for the fitting. This results in 11 degrees being used for the fit. 
\end{enumerate}

 We run both the cases for the reduced sample set, i.e. without the TRAPPIST-1 system, to decouple the influence of said system, from that of a Bernstein polynomial M-R fit with a reduced number of degrees. The resultant fits for these are shown in \autoref{fig:Trappist_impact} for comparison with \autoref{fig:results}.}

 \begin{figure*}[!t] 
\centering
\includegraphics[width=\textwidth]
{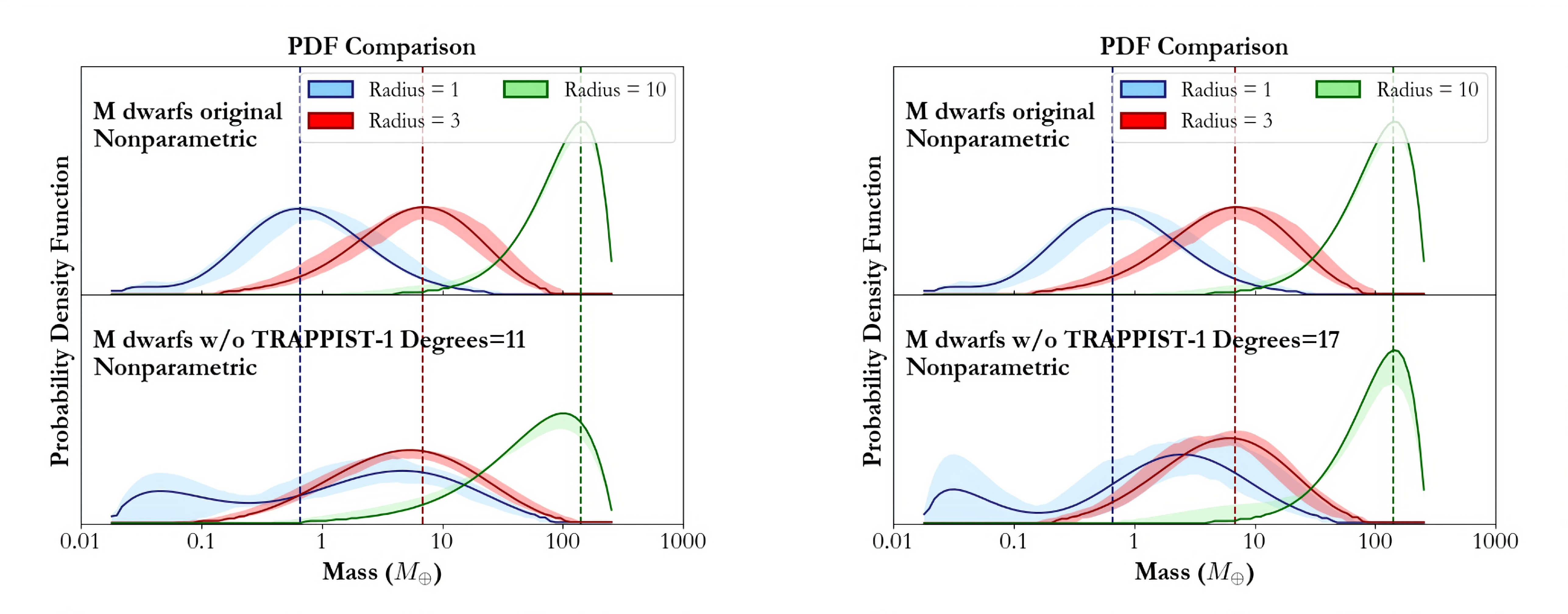}
\caption{Similar to \autoref{fig:pdf_compare}, we compare the CPDF of the M dwarf sample set with and without the TRAPPIST-1 system. The left plot shows the comparison for the reduced sample set, with 11 degrees used to fit the relationship. The right plot uses 17 degrees for the reduced sample set.  For both the cases, we notice that the 1 $R_{\oplus}$ CPDF is unconstrained in the absence of the TRAPPIST-1 system. The 3 and 10 $R_{\oplus}$ density functions are altered in the left plot and not the right. This suggests that the reduced number of degrees (11 vs. 17) being used to fit the joint distribution is smoothening the CPDFs.} \label{fig:trappist_pdf_compare} 
\end{figure*}
 
 \added{
 After fitting this relationship, we look at the resultant probability distribution function that we obtain when we marginalize this distribution to use it as a predictive function. Similar to \autoref{fig:pdf_compare}, we compare the predicting functions without TRAPPIST-1 in \autoref{fig:trappist_pdf_compare}.}

\section{Impact of sample size on nonparametric methods}\label{sec:simulation}

As discussed in \S \ref{sec:fitting}, we run a simulation to test the effectiveness of the nonparametric method as a fitting and predictive tool.  In particular, we visually assess the ability of the Bernstein polynomial model to qualitatively reproduce a known underlying distribution as a function of sample size and astrophysical scatter. We set the known distribution to be a power law in M-R space with a slope of $\sim$ 2.3 (mass as a function of radius). We simulated synthetic datasets from this known distribution and added mass and radius uncertainties of 10$\%$. We then tested this dataset for three different values of intrinsic dispersion, i.e. the scatter of the data points around the original power law. The tested values of intrinsic dispersion are 0, 0.1, and 0.5 in units of log($m$). 

\begin{figure*}[] 
\centering
\includegraphics[width=\textwidth]
{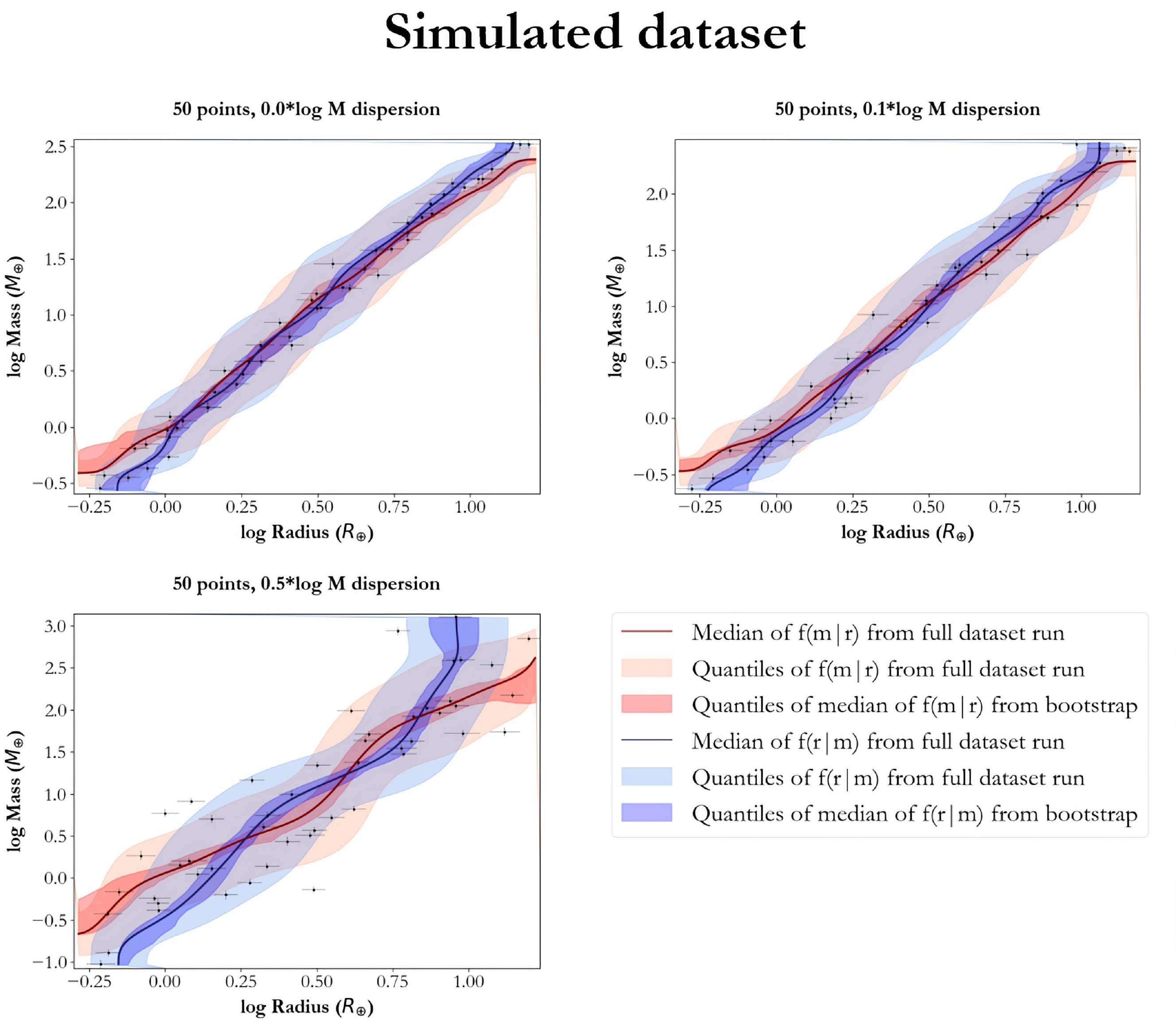}
\caption{Nonparametric fits to a simulated sample of 50 planets with an increasing amount of dispersion around the power law. The nonparametric method reproduces the power law well when the intrinsic dispersion is low. } \label{fig:sim_disp}
\end{figure*}

\begin{figure*}[] 
\centering
\includegraphics[width=\textwidth]
{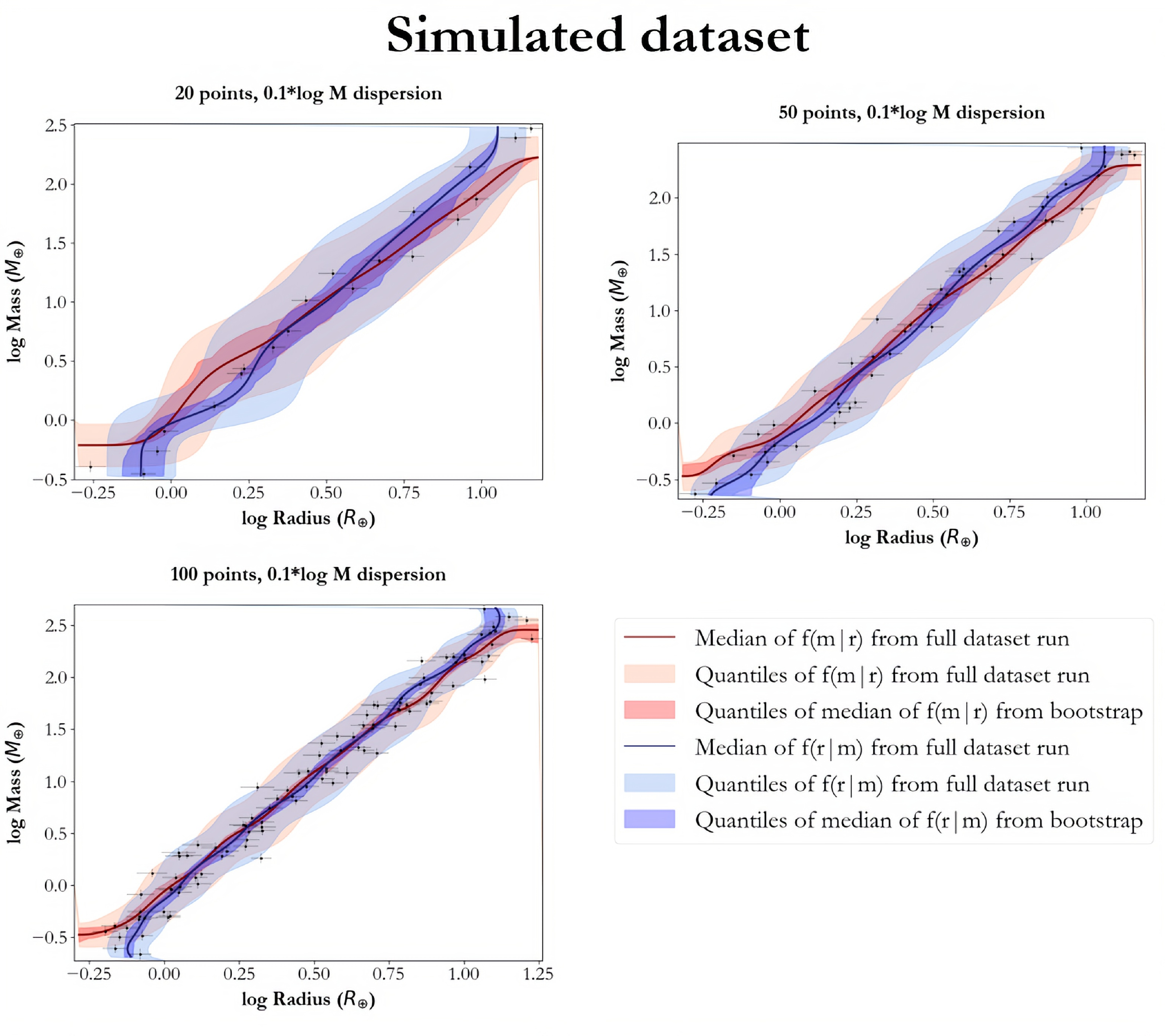}
\caption{Nonparametric fit to simulated samples of 20, 50, and 100 planets with the same amount of intrinsic scatter.  The fit improves as the sample size increases.  One can also note that the conditional f$(m|r)$ and f$(r|m)$ distributions begin to converge as the sample size increases.} \label{fig:sim_size}
\end{figure*}

We note that in the simulation, the nonparametric algorithm fits the simulated dataset well in the case of low intrinsic dispersion, even for small datasets ($\sim$ 20 points; see \autoref{fig:sim_size}). However, as we increase the intrinsic dispersion, the fit compensates by increased uncertainties and visual departures from the known underlying distribution (see \autoref{fig:sim_disp}).  \added{ Our simulation demonstrates that the nonparametric technique can reproduce the underlying (power law in this test case) distribution with as little as 20 points, with its precision improving as the number of points grows (\autoref{fig:sim_size})} \deleted{Therefore, this simulation confirms that the utility of the nonparametric method increases as the sample set increases in size (\autoref{fig:sim_size}). Importantly, we note that this simulation uses a simple power law, for which a parametric fit would be a sufficient descriptor of the data. However, we do not know the true underlying distribution of nature's exoplanet masses and radii, and so we are justified in applying a nonparametric model to real data with a large enough sample size.} 

\section{Discussion}\label{sec:discussion}

\subsection{Nonparametric vs Parametric methods}
In \S \ref{sec:com_p_np}, we compare the results of the nonparametric and parametric fit to the M dwarf sample. A parametric model is useful for small data sets, where it offers an easy means to develop an intuitive understanding for the data. It can also be used when there is a known physical process that is being fit or tested for. However, in cases where we need to explore the relationship and develop a forecasting or predicting routine, nonparametric methods offer a more flexible and unbiased option. Furthermore, in cases where the underlying distributions of the variables are unknown, or the data could have small-scale structure that can not be captured by a power law, the nonparametric method is a better predictor. For this M dwarf sample, the parametric and nonparametric methods offer similar median predictions in the realm of limited sample size (see \autoref{fig:pdf_compare}). However, as has been shown by \cite{ning_predicting_2018} with the \textit{Kepler} sample, a simple parametric power law (or broken power law) does not necessarily capture the features in the data as the sample size grows. A parametric predicting method will not necessarily reflect the data with the same accuracy and has the potential for higher bias, meaning that the predicted masses or radii will be farther from the true masses or radii.  \added{Furthermore, a broken power law gives a disjoint intrinsic scatter across the regions, as well as making it difficult to predict values close to the transitions between different regions}. Therefore, we propose this nonparametric method for the M-R relationship and offer the \texttt{Python} package \texttt{MRExo} as a practical tool for making these predictions.

\subsection{M dwarf vs \textit{Kepler}}

\citet{neil_host_2018} fit an  M-R relationship on a sample of M dwarf exoplanets and compared it to a sample of exoplanets orbiting FGK dwarfs to try to find evidence for any dependence on the host star mass. They did so by expanding upon the HBM approach introduced in \citet{wolfgang_probabilistic_2016}, to include another dimension that accounts for a possible host star mass dependence. As their M dwarf sample consisted of only six planets with radii and mass measurements, they found that their M-R power law was slightly shallower than that for their FGK sample, yet still consistent with no host star mass dependence. Our sample size of 24 planets agrees with this when fit with a similar parametric relationship. 

On the other hand, we find in \S \ref{sec:compare_m_kepler} that the \textit{Kepler} and M dwarf predictive mass distributions are different for the smallest and largest planets, in the sense that the smallest and largest planets are, on average, less massive around M dwarfs than around Sun-like stars.  This result holds for both the nonparametric and the parametric M dwarf fit.  The difference appears in these comparisons but not in the parametric comparison. This is because, as shown by \cite{ning_predicting_2018} the parametric model is not flexible enough to fit the larger \textit{Kepler} data set, and hence does not provide a reasonable baseline for comparison.

Astrophysically, this could indicate that the low-mass protoplanetary disks around M dwarfs result in lower-mass gas giants, which would explain the lower predictive masses for the 10 $R_{\oplus}$. \added{The HARPS M dwarf survey results \citep{bonfils_harps_2013}, along with other results from \citet{johnson_giant_2010} and \citet{gaidos_m_2014} point to a difference in mass distributions for M dwarfs vs FGK stellar-host planets for giant planets ($R \sim 10$ $R_{\oplus}$). The M dwarf giant planet (10 $R_{\oplus}$) occurrence rate seems to be fundamentally different from that of FGK stars. Therefore, we do not think that the difference in CPDFs for the 10 $R_{\oplus}$ is solely because of the fewer M dwarf giant planets, as compared to the \textit{Kepler} sample.}

For the smaller planets that are likely terrestrial, a lower mass at a given radius means a higher silicon-to-iron ratio, which may suggest that M dwarf disks process their refractory elements differently than Sun-like protoplanetary disks. 
\added{The small planet (1 $R_{\oplus}$) regime does suffer from a paucity of planets in the \textit{Kepler} sample. This can be attributed to an observational bias due to the difficulty in finding terrestrial planets around Sun like stars due to the smaller exoplanet detection signature. That being said, previous studies, such as \citet{mulders_increase_2015} have shown that there is a difference in planet occurrence rate for Earth-like planets around M stars versus FGK stars.}

Interestingly, the predictive mass distributions for 3 $R_{\oplus}$ planets are similar between M and FGK dwarf hosts.  This could indicate that the process of sub-Neptune formation is independent of the mass of the protoplanetary disk.  That said, the predictive mass distributions between the M dwarf and FGK planetary samples do overlap substantially, and the M dwarf dataset is still quite small and is dominated by the TRAPPIST-1 planets.  Therefore, these preliminary results should be revisited once more transiting M dwarf planets are discovered and followed up with ground-based observations.

\added{We also note that the orbital period ranges for the two samples differ. For the \textit{Kepler} sample, the orbital periods range from about 0.3 to 1100 days, whereas for the M dwarf samples, they vary from 1.5 to about 33 days. Therefore, future work that incorporates the orbital period dimension into a higher dimension relation could lead to further detailed insights into the orbital period dependence of exoplanet M-R relations around different exoplanet hosts. It would be very interesting to see how these results hold in this higher dimensional space as more transiting M dwarf planets are discovered and have their masses measured.}

\added{\subsection{Estimating the impact of the TRAPPIST-1 system on  M dwarf M-R relationship} In $\S$ \ref{sec:wo_Trappist}, we study the impact of the TRAPPIST-1 system on our M dwarf M-R relationship. In the $<$1.2 $R_{\oplus}$ regime, we have eight planets, of which seven are from said system, and the last one is \textit{Kepler}-138b. Even though no studies have so far suggested that the TRAPPIST-1 system is unusual in any aspect, we entertain the possibility by conducting a check and comparing the results with and without the TRAPPIST-1 system. However, by removing seven of the eight planets in that region, the prediction is unconstrained (Fig \ref{fig:trappist_pdf_compare}). Therefore, at this point, we cannot conclude if the difference between the M dwarf and \textit{Kepler} sample predictions for the 1 $R_{\oplus}$ predictions is because of a peculiarity in the TRAPPIST-1 system. However, we are very optimistic that this will be further probed with mass measurements of \textit{TESS} discoveries of transiting Earth-like planets around M dwarfs.

}

\subsection{Future Prospects}
As was seen with the \textit{Kepler} sample, as the sample size increases, we can start to unearth interesting phenomena that would be hard to discern from small data sets. In addition, more precise measurements with smaller error bars would help in more accurate M-R model fits. This would further unveil features and regions that were indistinguishable earlier. This will particularly hold true as the M dwarf M-R space starts to fill up with radii from \textit{TESS} \citep[]{ricker_transiting_2014}, and the advent of high-precision RV follow-up instruments like HPF \citep[]{mahadevan_habitable-zone_2012}, NEID \citep[]{schwab_design_2016}, MAROON-X \citep{seifahrt_development_2016}, MINERVA-RED \citep{blake_minerva-red:_2015}, CARMENES \citep{quirrenbach_carmenes:_2016}, SPIROU \citep{thibault_spirou_2012}, IRD \citep{kotani_infrared_2018}, ESPRESSO \citep{hernandez_espresso_2017}, NIRPS \citep{wildi_nirps:_2017}, SPIRou \citep{artigau_spirou:_2014}, iSHELL \citep{cale_precise_2018}, GIANO \citep{claudi_giarpstng:_2017} and EXPRES \citep{evans_expres:_2016}.

\section{Conclusion}\label{sec:conclusion}

In this paper, we fit the M-R relationship for a sample of 24 exoplanets around M dwarfs using nonparametric and parametric methods. Considering the small sample size, the measurements are currently well described by a power law, which we fit with a parametric hierarchical Bayesian model. We then discuss the deficiencies of parametric methods and the utility of nonparametric models, especially as the sample size increases. To further illustrate this point, we also run a simulation study showing how the nonparametric fit changes with a change in sample size and dispersion in the sample. We then compare the nonparametric and parametric results, finding them to be similar, on average, but currently with a larger variance for the nonparametric mass predictions. We also discuss differences in the M-R relationship for M dwarf versus \textit{Kepler} exoplanets and note that the predicted conditional probability density functions differ for the smallest and largest planets. Using this comparison, we illustrate the utility of an M dwarf M-R relationship in an era of the exciting new discoveries with \textit{TESS} and ground-based precision RV instrumentation.

We also introduce a new \texttt{Python} package called \texttt{MRExo}, which can be used as a predictive tool, as well as to fit nonparametric M-R relationships to new datasets. This code is available to the community to use in its own applications.

\section{Acknowledgements}
\added{The authors also thank the referee for providing insightful comments, that improved this work and made the results clearer.}
This research has made use of the NASA Exoplanet Archive, which is operated by the California Institute of Technology, under contract with the National Aeronautics and Space Administration under the Exoplanet Exploration Program. This paper includes data collected by the \textit{Kepler} mission. Funding for the \textit{Kepler} mission is provided by the NASA Science Mission directorate. This research has made use of NASA's Astrophysics Data System Bibliographic Services. Computations for this research were performed on the Pennsylvania State University's Institute for CyberScience Advanced CyberInfrastructure (ICS-ACI). This work was partially supported by funding from the Center for Exoplanets and Habitable Worlds. The Center for Exoplanets and Habitable Worlds is supported by the Pennsylvania State University, the Eberly College of Science, and the Pennsylvania Space Grant Consortium. S.K. acknowledges helpful discussions with Joe Ninan regarding the code, and Michael G. Scott and Tippy Toes for help with the project. G.K.S wishes to acknowledge support from NASA Headquarters under the NASA Earth and Space Science Fellowship Program grant NNX16AO28H. A.W. acknowledges support from the National Science Foundation Astronomy and Astrophysics Postdoctoral Fellowship program under award No. 1501440.  We acknowledge support from NSF grants AST-1006676, AST-1126413, AST-1310885, and AST-1517592 in our pursuit of precision radial velocities in the near-infrared to discover and characterize planets around M dwarfs.

\software{MRExo \citep{mrexo_2019},
            Astropy \citep{robitaille_astropy:_2013}, Numpy \citep{numpy},
          Matplotlib \citep{hunter_matplotlib:_2007},
           SciPy \citep{oliphant_python_2007}, iPython \citep{PER-GRA:2007}, Enthought Canopy Python. }

\appendix

\setcounter{table}{0}
\renewcommand{\thetable}{A\arabic{table}}

\begingroup 
 \renewcommand{\arraystretch}{0.3} 
\begin{table*}[]
\caption{Masses and Radii of M dwarf planets used in this work.}\label{tab:data}
\small
\begin{tabular}{l|l|l|l|l|l|l}

\toprule
Planet   & Mass     & $\sigma_M$ & Radius & $\sigma_R$ & Mass               & Radius     \\
Name & ($M_{\oplus}$) & ($M_{\oplus}$) & ($R_{\oplus}$) & ($R_{\oplus}$) & Reference & Reference \\ \midrule 
             &          &            &        &            &                             &                             \\
HATS-6 b     & 101.3878 & 22.2481    & 11.187 & 0.213      & \citet{hartman_hats-6b:_2015}         &  \citet{hartman_hats-6b:_2015}        \\
             &          &            &        &            &                             &                             \\
GJ 1214 b    & 6.26125  & 0.85814    & 2.847  & 0.202      &  \citet{harpsoe_transiting_2013} & \citet{harpsoe_transiting_2013} \\
             &          &            &        &            &                             &                             \\
K2-22 b$^{\rm\ref{UpperLim}}$       & 444.962  & 0          & 2.5    & 0.4        & \citet{sanchis-ojeda_k2-esprint_2015}   & \citet{sanchis-ojeda_k2-esprint_2015}  \\
             &          &            &        &            &                             &                             \\
K2-18 b      & 7.96     & 1.91       & 2.38   & 0.22       & \citet{cloutier_characterization_2017}        & \citet{cloutier_characterization_2017}       \\
             &          &            &        &            &                             &                             \\
LHS 1140 b   & 6.98     & 0.89       & 1.727  & 0.032      & \citet{ment_second_2018}           & \citet{ment_second_2018}          \\
             &          &            &        &            &                             &                             \\
NGTS-1 b     & 258.078  & 22.40702   & 14.908 & 5.268      & \citet{bayliss_ngts-1b:_2018}         & \citet{bayliss_ngts-1b:_2018}        \\
             &          &            &        &            &                             &                             \\
K2-137 b\footnote{\label{UpperLim} Upper limits}     & 158.915  & 0          & 0.89   & 0.09       & \citet{smith_k2-137_2018}            & \citet{smith_k2-137_2018}          \\
             &          &            &        &            &                             &                             \\
K2-33 b$^{\rm\ref{UpperLim}}$      & 1175.971 & 0          & 5.04   & 0.355      & \citet{mann_zodiacal_2016}            & \citet{mann_zodiacal_2016}           \\
             &          &            &        &            &                             &                             \\
LHS 1140 c   & 1.81     & 0.39       & 1.282  & 0.024      & \citet{ment_second_2018}            & \citet{ment_second_2018}            \\
             &          &            &        &            &                             &                             \\
GJ 436 b     & 23.1     & 0.8        & 4.191  & 0.109      & \citet{turner_ground-based_2016}         & \citet{turner_ground-based_2016}          \\
             &          &            &        &            &                             &                             \\
GJ 1132 b    & 1.66     & 0.23       & 1.43   & 0.16       & \citet{bonfils_radial_2018}         & \citet{southworth_detection_2017}      \\
             &          &            &        &            &                             &                             \\
Kepler-32 b  & 9.4      & 3.35       & 2.231  & 0.072      & \citet{hadden_densities_2014}         & \citet{berger_revised_2018}          \\
             &          &            &        &            &                             &                             \\
Kepler-32 c  & 7.7      & 4.4        & 2.112  & 0.071      & \citet{hadden_densities_2014}          & \citet{berger_revised_2018}          \\
             &          &            &        &            &                             &                             \\
GJ 3470 b    & 13.9     & 1.5        & 4.57   & 0.18       & \citet{awiphan_transit_2016}        & \citet{awiphan_transit_2016}         \\
             &          &            &        &            &                             &                             \\
TRAPPIST-1 b & 1.017    & 0.1485     & 1.121  & 0.0315     & \citet{grimm_nature_2018}           & \citet{grimm_nature_2018}           \\
             &          &            &        &            &                             &                             \\
TRAPPIST-1 c & 1.156    & 0.1365     & 1.095  & 0.0305     & \citet{grimm_nature_2018}           & \citet{grimm_nature_2018}           \\
             &          &            &        &            &                             &                             \\
TRAPPIST-1 d & 0.297    & 0.037      & 0.784  & 0.023      & \citet{grimm_nature_2018}           & \citet{grimm_nature_2018}           \\
             &          &            &        &            &                             &                             \\
TRAPPIST-1 e & 0.772    & 0.077      & 0.91   & 0.0265     & \citet{grimm_nature_2018}           & \citet{grimm_nature_2018}           \\
             &          &            &        &            &                             &                             \\
TRAPPIST-1 f & 0.934    & 0.079      & 1.046  & 0.0295     & \citet{grimm_nature_2018}           & \citet{grimm_nature_2018}           \\
             &          &            &        &            &                             &                             \\
TRAPPIST-1 g & 1.148    & 0.0965     & 1.148  & 0.0325     & \citet{grimm_nature_2018}           & \citet{grimm_nature_2018}           \\
             &          &            &        &            &                             &                             \\
TRAPPIST-1 h & 0.331    & 0.0525     & 0.773  & 0.0265     & \citet{grimm_nature_2018}           & \citet{grimm_nature_2018}           \\
             &          &            &        &            &                             &                             \\
Kepler-45 b  & 159      & 19         & 11.634 & 0.362      & \citet{southworth_homogeneous_2012}      & \citet{berger_revised_2018}          \\
             &          &            &        &            &                             &                             \\
Kepler-138 b & 0.066    & 0.048      & 0.629  & 0.0275     & \citet{jontof-hutter_mass_2015}   & \citet{berger_revised_2018}          \\
             &          &            &        &            &                             &                             \\
Kepler-138 c & 1.97     & 1.516      & 1.519  & 0.0975     & \citet{jontof-hutter_mass_2015}   & \citet{berger_revised_2018}          \\
             &          &            &        &            &                             &                             \\
Kepler-138 d & 0.64     & 0.5305     & 1.323  & 0.043      & \citet{jontof-hutter_mass_2015}   & \citet{berger_revised_2018}          \\
             &          &            &        &            &                             &                             \\
Kepler-26 b  & 5.02     & 0.66       & 3.191  & 0.095      &  \citet{hadden_kepler_2017} & \citet{berger_revised_2018}          \\
             &          &            &        &            &                             &                             \\
Kepler-26 c  & 6.12     & 0.71       & 2.976  & 0.253      & \citet{hadden_kepler_2017}   & \citet{berger_revised_2018}  \\ \hline       
\end{tabular}
\end{table*}
\endgroup

\bibliography{MRExo}
\renewcommand*{\bibfont}{\footnotesize}

\listofchanges

\end{document}